\documentclass{article}

\usepackage{arxiv}

\usepackage[utf8]{inputenc} 
\usepackage[T1]{fontenc}    
\usepackage{hyperref}       
\usepackage{url}            
\usepackage{booktabs}       
\usepackage{amsfonts}       
\usepackage{nicefrac}       
\usepackage{microtype}      
\usepackage{lipsum}
\usepackage{graphicx}
\usepackage{algorithmic}
\usepackage[tight,footnotesize]{subfigure}
\usepackage{caption}
\usepackage{adjustbox}
\usepackage{makecell} 
\usepackage[linesnumbered,ruled]{algorithm2e}

\usepackage{pifont}

\usepackage{authblk}  

\graphicspath{ {./images/} }

\title{Agentsway --- Software Development Methodology for AI Agents-based Teams}

\author[1]{Eranga Bandara}
\author[1]{Ross Gore}
\author[3]{Xueping Liang}
\author[7]{Sachini Rajapakse}
\author[7]{Isurunima Kularathna}
\author[10]{Pramoda Karunarathna}
\author[1]{Peter Foytik}
\author[1]{Sachin Shetty}
\author[1]{Ravi Mukkamala}
\author[2]{Abdul Rahman}
\author[6]{Amin Hass}
\author[4]{Ng Wee Keong}
\author[5]{Kasun De Zoysa}
\author[8]{Aruna Withanage}
\author[8]{Nilaan Loganathan}

\affil[1]{Old Dominion University, USA \\
\texttt{\{cmedawer, rgore, pfoytik, sshetty, mukka\}@odu.edu}}

\affil[2]{Deloitte \& Touche LLP, USA \\
\texttt{abdulrahman@deloitte.com}}

\affil[3]{Florida International University, USA \\
\texttt{xuliang@fiu.edu}}

\affil[4]{Nanyang Technological University, Singapore \\
\texttt{awkng@ntu.edu.sg}}

\affil[6]{AnaletIQ, VA, USA \\
\texttt{aminhass@analetiq.com}}

\affil[7]{IcicleLabs.AI \\
\texttt{\{sachini.rajapakse, isurunima.kularathna\}@iciclelabs.ai}}

\affil[5]{University of Colombo, Sri Lanka \\
\texttt{kasun@ucsc.cmb.ac.lk}}

\affil[10]{University of Sri Jayewardenepura, Sri Lanka \\
\texttt{pramodabhavani@sjp.ac.lk}}

\affil[8]{Effectz.AI \\
\texttt{\{aruna, nilaan\}@effectz.ai}}

\begin{document}
\maketitle
\begin{abstract}
The emergence of Agentic AI is fundamentally transforming how software is designed, developed, and maintained. Traditional software development methodologies such as Agile, Kanban, ShapeUp, etc, were originally designed for human-centric teams and are increasingly inadequate in environments where autonomous AI agents contribute to planning, coding, testing, and continuous learning. To address this methodological gap, we present ``Agentsway" a novel software development framework designed for ecosystems where AI agents operate as first-class collaborators. Agentsway introduces a structured lifecycle centered on human orchestration, governance, and privacy-preserving collaboration among specialized AI agents. The framework defines distinct roles for planning, prompting, coding, testing, and fine-tuning agents, each contributing to iterative improvement and adaptive learning throughout the development process. By integrating fine-tuned Large Language Models (LLMs) that leverage outputs and feedback from different agents throughout the development cycle as part of a retrospective learning process, Agentsway enhances domain-specific reasoning, adaptive learning, and explainable decision-making across the entire software development lifecycle. Responsible AI principles are further embedded across the agents through the coordinated use of multiple fine-tuned LLMs and advanced reasoning models, ensuring balanced, transparent, and accountable decision-making. This work advances software engineering theory and practice by formalizing agent-centric collaboration, incorporating governance and privacy-by-design principles, and introducing measurable metrics for productivity, reliability, and trust. Agentsway represents a foundational step toward the next generation of AI-native, self-improving software development methodologies. To the best of our knowledge, this is the first research effort to introduce a dedicated methodology explicitly designed for AI agent–based software engineering teams. While this paper primarily focuses on managing AI agents within software development projects, the underlying principles can be readily extended to other collaborative, team-based environments that integrate AI agents.
\end{abstract}


\section{Introduction}

Software engineering is entering a transformative phase driven by the emergence of Agentic AI~\cite{agentic-ai}. Recent advances in Large Language Models (LLMs)~\cite{llm}, Vision-Language Models (VLMs)~\cite{vision-language-model, vistion-language-model-comparison}, and autonomous reasoning systems have empowered AI agents to perform tasks that were once exclusively carried out by humans, including requirements analysis, system design, implementation, testing, and deployment. This transformation challenges the foundational assumptions of traditional software development methodologies such as Agile, Kanban, Waterfall, ShapeUp, and Extreme Programming (XP)~\cite{agile-information-journal, kanban, shapeup, xp}. These methods were conceived for human-only collaboration, manual iteration cycles, and team-driven decision-making processes~\cite{sdm-comparison, sdm}.

In this emerging Agentic Software Era~\cite{agentic-ai}, development ecosystems are increasingly composed of both humans and autonomous AI agents that operate at digital speed, handle complex reasoning chains, and adapt continuously to dynamic contexts. Such hybrid environments demand new methodological foundations centered on autonomy, orchestration, and explainability~\cite{sd-agents}. While significant progress has been made in AI-assisted programming, agentic workflow design, and multi-agent coordination frameworks, there remains a critical methodological gap: the lack of a formal process model that defines how AI agents should systematically collaborate—with each other and with human supervisors—throughout the entire software development lifecycle.

To address this gap, we propose ``Agentsway", a novel software development methodology explicitly designed for AI-agent–centric environments. Agentsway rethinks the traditional team structure by positioning the human primarily as an orchestrator while delegating specialized functions to AI agents. These include \textit{Planning Agents}, \textit{Prompting Agents}, \textit{Coding Agents}, \textit{Testing Agents}, and \textit{Fine-Tuning Agents}. Each role contributes to a cyclical, privacy-preserving, and self-improving lifecycle governed by transparent interaction rules and human oversight.


The framework integrates fine-tuned LLMs to enhance domain-specific reasoning, adaptive learning, and explainable decision-making. A core feature of Agentsway is its privacy-by-design architecture, which guarantees that model training, fine-tuning, and data exchange occur entirely within secure organizational and regulatory boundaries. Furthermore, responsible AI principles~\cite{responsbile-ai-llm} are embedded across the agents through the integration of multiple fine-tuned LLMs with advanced reasoning LLM~\cite{reasoning-llms, bassa-llama}. Collectively, these design principles establish Agentsway as a structured, ethical, and scalable foundation for AI-native software engineering in the era of autonomous systems. The primary contributions of this work are summarized as follows:

\begin{enumerate}
\item Proposed a novel software development methodology, ``Agentsway", designed specifically for AI agent–based teams.
\item Introduced a human-in-the-loop orchestration model, where the human acts as the primary orchestrator while delegating specialized functions—planning, prompting, coding, testing, and LLM fine-tuning—to autonomous AI agents.
\item Developed a continuous improvement mechanism for LLM fine-tuning that leverages outputs and feedback from different agents throughout the development cycle as part of a retrospective learning process, ensuring both model refinement and adherence to responsible AI principles.
\item Demonstrated the applicability and usability of the Agentsway methodology through a real-world use case focused on an agentic AI workflow for legal case handling automation.
\end{enumerate}

The remainder of this paper is organized as follows. Section 2 introduces the agent team structure within the Agentsway methodology and explains the distinct roles of human and AI agents. Section 3 details the core functionalities of the Agentsway framework, describing how each component contributes to the overall software development lifecycle. Section 4 presents a real-world use case—legal case handling automation—implemented with Agentsway method. Section 5 discusses the evaluation of key components, including the Planning Agent, Prompting Agent, and Fine-Tuning process. Finally, Section 6 concludes the paper and outlines directions for future research and development.

\section{Entities and Agents Defined in the Agentsway}

\begin{figure}[h]
\centering{}
\includegraphics[width=5.2in]{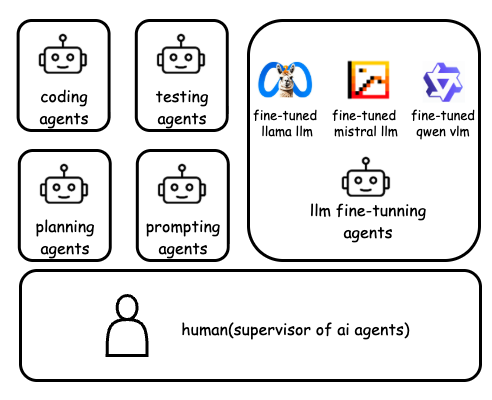}
\DeclareGraphicsExtensions.
\vspace{-0.2in}
\caption{Agents defined in the Agentsway methodology.}
\label{agentika-architecture}
\end{figure}

\begin{figure}[h]
\centering{}
\includegraphics[width=5.2in]{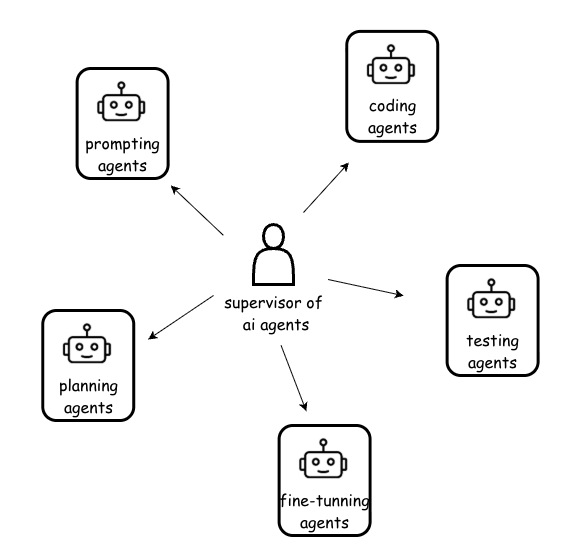}
\DeclareGraphicsExtensions.
\vspace{-0.2in}
\caption{Developer surrounded by multiple AI Agents.}
\label{agentika-surrounded-agents}
\end{figure}

\label{sec:agents}

The Agentsway methodology defines a structured team architecture in which a single human orchestrator collaborates with a suite of specialized AI agents, as illustrated in Figure~\ref{agentika-architecture}. The human (e.g., developer) is surrounded by AI agents and supervises their interactions, as shown in Figure~\ref{agentika-surrounded-agents}. Each agent fulfills a specific role within the software development lifecycle, ensuring a clear division of labor, traceability, and continuous adaptation. In this setup, all core tasks are performed by AI agents, while the human primarily oversees, validates, and guides their activities. The overall process harmonizes human oversight with agentic autonomy, creating a self-improving workflow where insights from each development iteration are used to refine and optimize subsequent cycles through fine-tuned model updates. The following are the main entities and agents defined within the Agentsway framework.

\subsection{Human Orchestrator}

The human participant functions as the orchestrator of the entire development process. Humans are primarily responsible for interpreting high-level business goals, conducting stakeholder interactions, and ensuring that the final outputs align with organizational objectives. During requirements gathering—such as project meetings or planning sessions—AI agents assist in note-taking, document summarization, and context extraction, allowing the human to focus on conceptual and strategic aspects. The human validates artifacts generated by AI agents, such as task lists, pitches, or code outputs, and provides governance and ethical oversight across all stages of development.

\subsection{Planning Agent}

The Planning Agent acts as the central reasoning and coordination component in Agentsway. It examines all available project documents, meeting summaries, and contextual artifacts to understand requirements and decompose them into executable tasks. Leveraging fine-tuned LLMs~\cite{mistral-fine-tune, slice-gpt} trained on historical organizational data, the Planning Agent generates detailed task descriptions, resource estimates, and project pitches~\cite{shapeup}. These deliverables are published to collaborative platforms such as GitHub for version control and transparency. The human orchestrator then verifies and approves the generated plans before execution begins, ensuring consistency with strategic priorities.

\subsection{Prompting Agent}
The Prompting Agent serves as the bridge between planning and implementation. It analyzes each approved task and constructs detailed, context-aware prompts tailored for code generation by the Coding Agents. These prompts encapsulate functional requirements, coding style preferences, and integration dependencies. The agent relies on fine-tuned LLMs to synthesize optimal prompts based on prior project knowledge and team conventions. Generated prompts are published to GitHub or shared repositories for human review and refinement, maintaining a balance between automation and oversight.

\subsection{Coding Agents}
The Coding Agents (e.g., Claude Code, Codex, or similar LLM-based development assistants~\cite{claude-code, claude-codex}) are responsible for translating approved prompts into executable code. They operate within defined project environments, adhering to the coding standards and architectural constraints specified by the organization. Coding Agents collaborate autonomously to implement features, refactor existing modules, and document their outputs. Human oversight remains essential for validating implementation quality, ensuring maintainability, and performing high-level design reviews.

\subsection{Testing Agents}
The Testing Agents ensure the correctness, reliability, and security of the produced software. They execute automated unit, integration, and regression tests while also performing static analysis and vulnerability scans. Testing Agents produce structured reports highlighting defects, performance metrics, and coverage statistics. Human developers may complement this process with manual exploratory testing and verification, particularly for edge cases and usability aspects that require human judgment~\cite{reasoning-llms}. The integration of testing results into version control enables continuous quality monitoring throughout the development cycle.

\subsection{Fine-Tuning Agents}
The Fine-Tuning Agents constitute the learning and improvement layer of the Agentsway methodology. After each development cycle, these agents collect task data, prompts, generated code, and testing feedback to refine pre-trained LLMs(e.g., Llama-3, Pixtral, Qwen~\cite{llama-3, pixtral, qwen2}) through incremental fine-tuning. This retrospective process allows the system to improve contextual accuracy, adaptability, and compliance over time. By retraining within secure and privacy-preserving environments, Fine-Tuning Agents ensure that organizational knowledge and sensitive data remain protected while enhancing the system’s capability for future projects.

\section{Functionalities of the Agentsway}

\label{sec:functionalities}

The Agentsway framework defines a series of core functionalities that structure the complete software development lifecycle. Each functionality corresponds to a specific phase, supported by one or more specialized agents. These functions ensure that knowledge flows seamlessly between human oversight and autonomous execution, maintaining transparency, traceability, and continuous learning.

\subsection{Requirement Gathering and Human Interaction}

The initial phase involves understanding and formalizing project requirements through human–AI collaboration. The human orchestrator participates in meetings with stakeholders to capture business goals, constraints, and acceptance criteria~\cite{agile-information-journal}. During these interactions, AI note-taking agents automatically transcribe discussions, extract key points, and summarize decisions using natural language understanding techniques. They also gather supplementary documents such as specifications, emails, and reports for contextual grounding. This hybrid approach reduces cognitive load on the human participant while ensuring no critical detail is lost. The resulting artifacts are stored in a structured repository that later informs the planning phase.

\subsection{Planning}

In the planning functionality, the Planning Agent processes all collected documentation to develop a clear technical roadmap. Using fine-tuned LLMs trained on historical organizational data, the agent decomposes high-level requirements into actionable development tasks, defines dependencies, and estimates the complexity of each activity. It produces structured deliverables such as pitches, user stories, and task lists, which are then committed to collaborative repositories (e.g., GitHub)~\cite{shapeup}. The human orchestrator reviews these deliverables to validate their alignment with project priorities and business objectives before authorizing execution. This phase establishes the foundation for coherent downstream automation.

\subsection{Prompting}

Once planning is approved, the Prompting Agent transforms validated tasks into detailed and context-aware prompts suitable for code-generation models~\cite{claude-code}. This agent leverages prior data and domain-specific fine-tuning to craft optimal prompt structures that balance instruction clarity with creative freedom~\cite{prompt-engineering}. It may integrate information from previous iterations or fine-tuned models to ensure continuity and adherence to organizational standards. Generated prompts are logged and version-controlled for auditability. The human orchestrator verifies these prompts to confirm that they accurately capture the intended requirements and constraints.

\subsection{Coding}

The Coding Agents are responsible for translating approved prompts into executable software artifacts. They utilize state-of-the-art code generation models (e.g., Claude Code, Codex, or GPT-based systems)~\cite{claude-code, claude-codex} to produce source code, documentation, and integration scripts. These agents operate autonomously within sandboxed environments and interact with repositories, build systems, and APIs via the Model Context Protocol (MCP). Code is continuously tested, reviewed, and refined in response to agent feedback and human evaluation. This functionality significantly accelerates development cycles while maintaining human supervision to guarantee design integrity and security compliance.

\subsection{Testing}
The Testing Agents form the quality assurance layer of the Agentsway system. They perform automated unit, integration, and regression testing while monitoring performance, scalability, and security aspects of the software. The agents analyze logs, detect anomalies, and generate test coverage reports that guide both humans and coding agents in corrective actions~\cite{claude-code, claude-codex}. Humans participate in manual verification for usability and exploratory testing, particularly in scenarios requiring contextual judgment. By integrating with continuous integration/continuous deployment (CI/CD) pipelines, the testing functionality ensures real-time feedback loops and early defect detection.

\subsection{LLM Fine-Tuning}
\begin{figure}[h]
\centering{}
\includegraphics[width=5.2in]{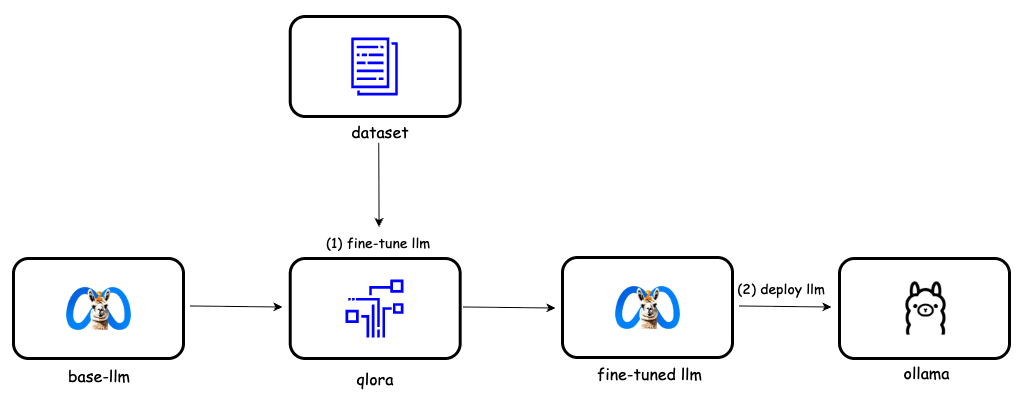}
\vspace{-0.1in}
\DeclareGraphicsExtensions.
\caption{Fine-tune LLMs with Qlora and deploy with Ollama.}
\label{llm-fine-tune}
\end{figure}

The final functionality of the Agentsway lifecycle is continuous learning through the Fine-Tuning Agents. After each development cycle, these agents collect task-related data—such as prompts, code diffs, testing feedback, and human annotations—and use it to fine-tune pre-trained LLMs(e,g Llama-3, Pixtral, Qwen~\cite{llama-3, pixtral, qwen2, devsec-gpt}), Figure~\ref{llm-fine-tune}. This process represents the system’s retrospective learning mechanism, where experience from completed tasks enhances future reasoning and performance. Fine-tuning operations occur within secure organizational boundaries, preserving data privacy and compliance with regulations such as GDPR. Over time, this cyclical refinement process strengthens model reliability, contextual alignment, and decision-making accuracy, enabling Agentsway to evolve into a self-improving software development ecosystem.

\subsection{Implementing Responsible AI Agents}

\begin{figure}[h]
\centering{}
\includegraphics[width=5.2in]{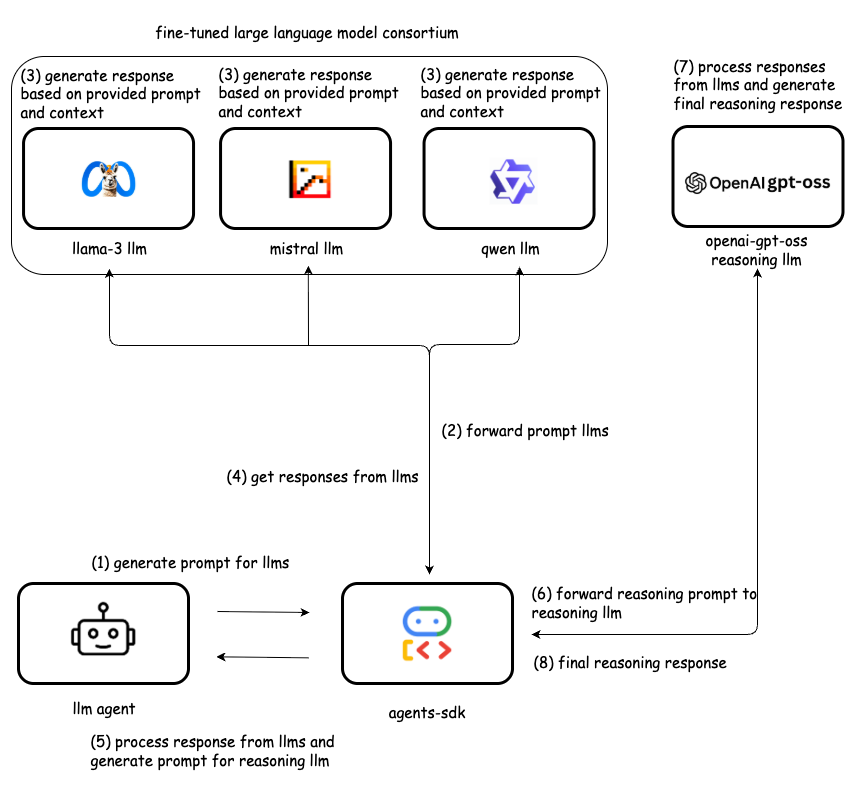}
\vspace{-0.1in}
\DeclareGraphicsExtensions.
\caption{LLM consortium and reasoning llm integration flow.}
\label{llm-consortium}
\end{figure}

The responsible AI aspects within the Agentsway framework are realized through the integration of multiple fine-tuned LLMs operating as a consortium, coupled with a dedicated reasoning LLM (e.g., OpenAI GPT-OSS)~\cite{reasoning-llms, gpt-oss}, as illustrated in Figure~\ref{llm-consortium}. Each agent (e.g., planning, prompting, or testing) can be configured to interface with this LLM consortium, ensuring balanced, multi-perspective reasoning. For instance, when executing a planning task, the agent generates prompts that are distributed across multiple fine-tuned LLMs—such as Llama-3, Pixtral, Qwen~\cite{llama-3, pixtral, qwen2} obtain diverse responses. These outputs are then processed by the reasoning LLM, which synthesizes and validates the inputs to produce a coherent and responsible final decision or artifact (e.g., a project pitch or structured plan). This ensemble-based reasoning mechanism enhances both the accuracy and accountability of agentic decisions, providing a robust foundation for responsible AI behavior across the software development lifecycle.

\section{Use Case of the Agentsway}

To demonstrate the applicability of Agentsway, we present a use case focused on legal case handling automation. The objective is to automate document retrieval, case summarization, and legal question answering across large-scale corpora, while preserving human oversight and data privacy. The system integrates the OpenAI Agents SDK for orchestrating LLM-based agents, Qdrant as the vector store for semantic document retrieval, and Mem0 for conversational memory management, with interoperability provided through the MCP.

The process begins with requirement capture and planning. The human orchestrator collaborates with legal professionals to define key objectives such as summarizing cases by file name or full-text input, and answering precedent-related questions. The Planning Agent analyzes these requirements, decomposes them into executable flows, and generates structured pitches outlining dependencies, task breakdowns, and acceptance criteria. Among the defined flows are: the QAAssistantFlow for context-aware Q\&A, the SummarizationFlow for retrieving and summarizing cases by file name, the AttachedCaseSummarizationFlow for direct text summarization, and a Workflow Manager that performs intent classification and flow routing. These agent-generated plans are version-controlled and reviewed for legal and technical accuracy.

The Prompting, Coding, Testing, and Fine-Tuning Agents collectively execute, validate, and improve the workflow. The Prompting Agent transforms approved plans into flow-specific prompts that guide Coding Agents in building, retrieving, and summarizing case data through the Cluade code. Testing Agents ensure factual accuracy, style consistency, and compliance, while Fine-Tuning Agents aggregate results and feedback for continuous model refinement using frameworks such as Unsloth~\cite{llamafactory-unsloth, metavese-llama}. All learning and inference are performed within secure environments to ensure privacy, transparency, and accountability. Through this use case, Agentsway demonstrates its potential to enable scalable, trustworthy, and privacy-preserving agentic automation for complex, regulation-sensitive domains like law.

\begin{figure}[h]
\centering{}
\includegraphics[width=5.2in]{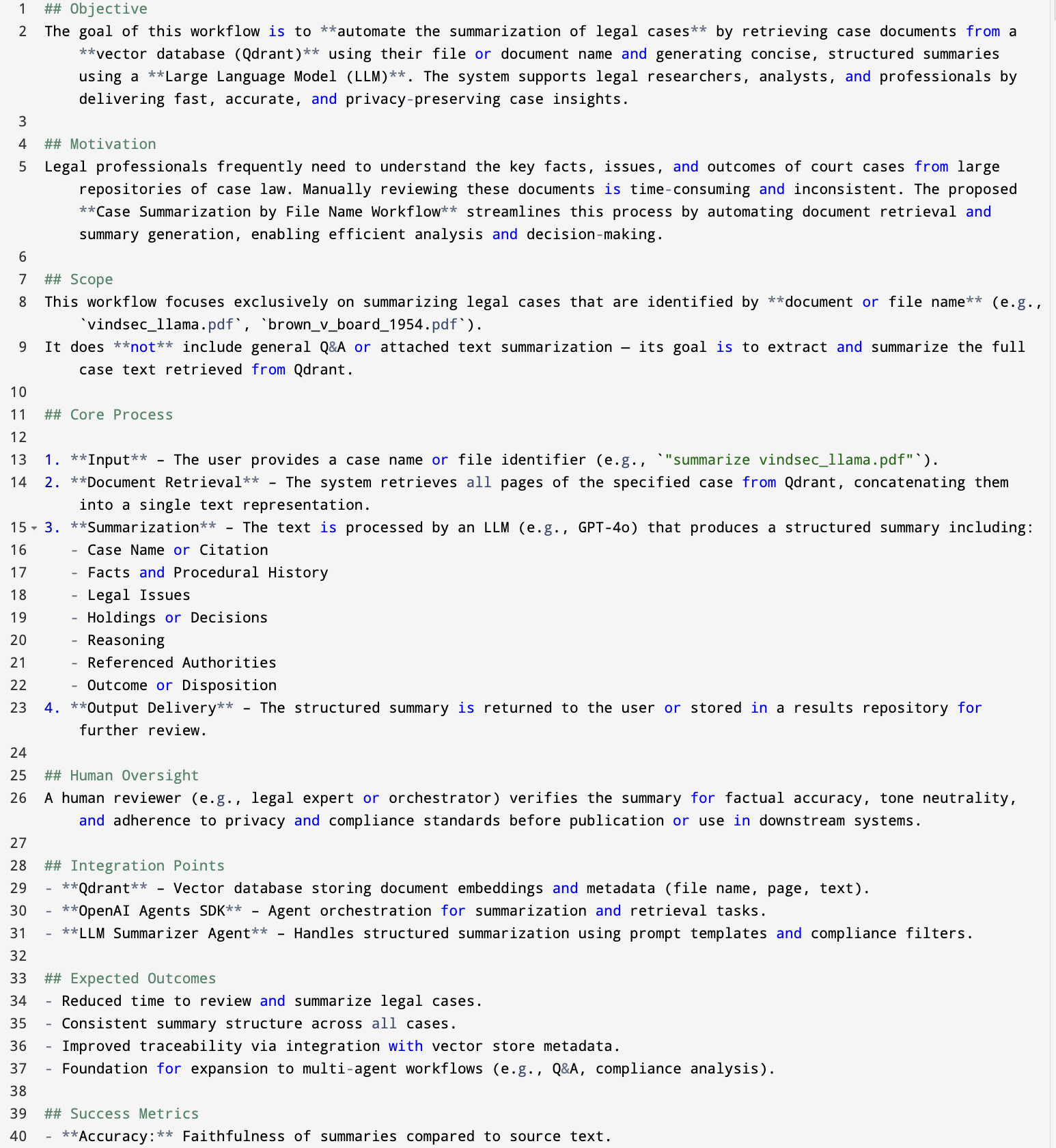}
\DeclareGraphicsExtensions.
\caption{Pitch generated by planning agent.}
\label{agentika-pitch}
\end{figure}

\begin{figure}[h]
\centering{}
\includegraphics[width=5.2in]{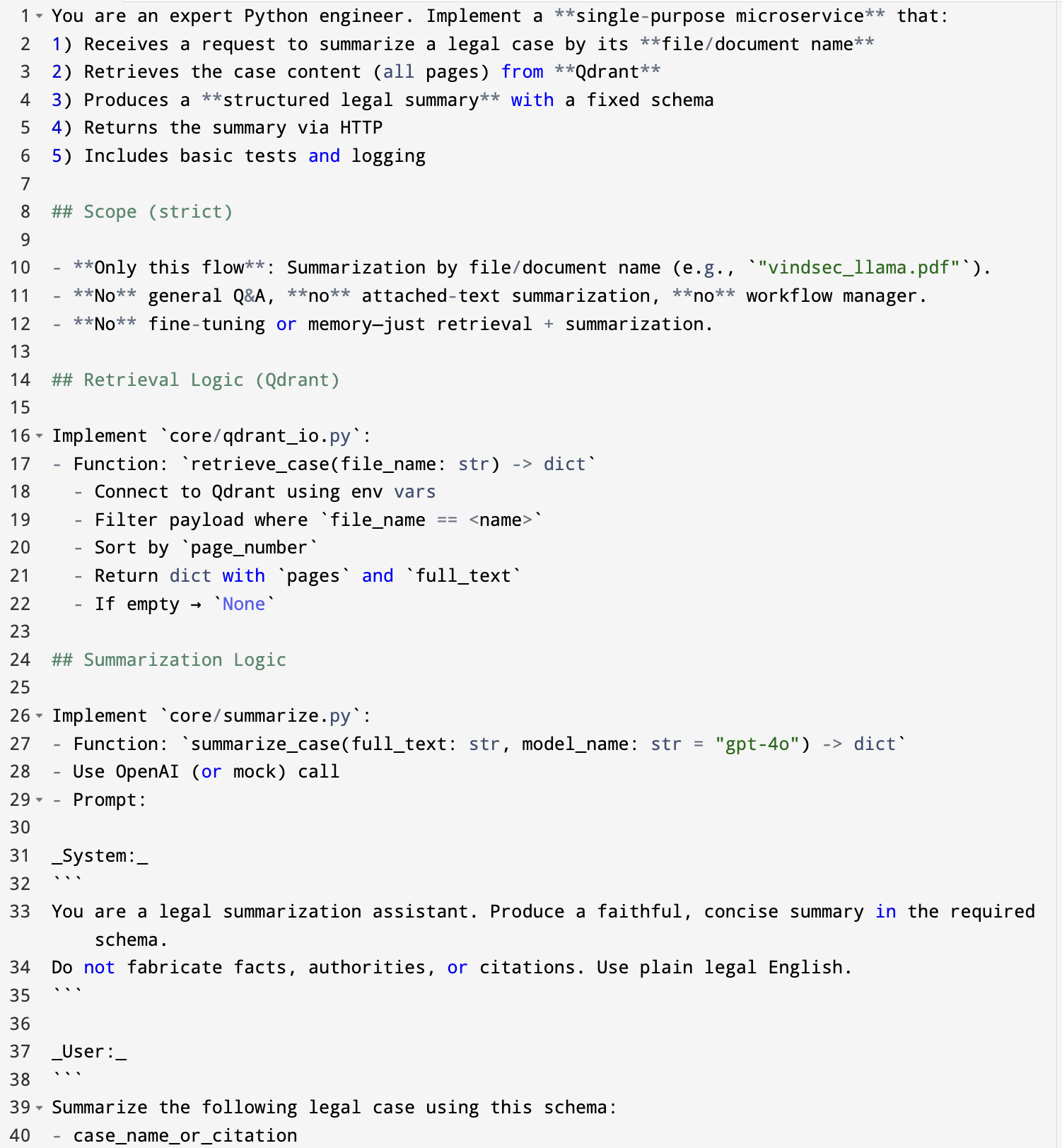}
\DeclareGraphicsExtensions.
\caption{Prompt generated by prompting agent.}
\label{agentika-prompt}
\end{figure}

\section{Evaluation}

\label{sec:evaluation}

The evaluation of the Agentsway methodology focuses on measuring the performance and reliability of its key functional components: the \textbf{Planning Agent}, the \textbf{Prompting Agent}, and the \textbf{Fine-Tuning process}. Each component was assessed in the context of the legal case handling use case, emphasizing the system’s ability to generate coherent plans, contextually accurate prompts, and demonstrable improvements through iterative fine-tuning. Evaluation criteria include quality, consistency, interpretability, and efficiency.

\paragraph{Evaluation of the Planning Agent.}

The Planning Agent was evaluated based on its capacity to analyze high-level requirements and autonomously generate structured task pitches — in this case, the Case Summarization Workflow Pitch. Starting from an abstract goal (“summarize legal cases retrieved from a vector store”), the agent successfully decomposed the requirement into actionable tasks such as document retrieval, summarization, output structuring, and human validation. The generated pitch clearly articulated objectives, scope, system components, and evaluation criteria in a format suitable for human and agent execution. As presented in the generate pitch, Figure~\ref{agentika-pitch}(full pitch available in githhub~\cite{erangaeb2025casesummary}), the Planning Agent demonstrated strong contextual understanding by correctly identifying dependencies (e.g., ``retrieve case → summarize → validate") and aligning them with the Agentsway framework’s privacy-preserving and modular principles. On a 5-point Likert scale, the generated pitch received an average rating of 4.7 for coherence, correctness, and implementation readiness. These results indicate that the Planning Agent can reliably transform abstract development goals into structured, executable workflow definitions with minimal human refinement.

\paragraph{Evaluation of the Prompting Agent.}

The Prompting Agent was evaluated based on its ability to autonomously generate precise, implementation-ready prompts for downstream coding agents—specifically for the Case Summarization by File Name workflow. Given the pitch produced by the Planning Agent, the Prompting Agent generated a detailed Claude Code prompt~\cite{prompt-engineering, vindsec-llams} that clearly defined objectives, scope, architecture, and acceptance criteria. The produced prompt, Figure~\ref{agentika-prompt}(full prompt available in github ~\cite{erangaeb2025claudeprompt}) outlined every critical aspect of development, including endpoint design, schema definitions, retrieval logic from Qdrant, and summarization behavior using an LLM, while explicitly excluding out-of-scope elements such as Q\&A or attached-text summarization.


\paragraph{Evaluation of LLM Fine-Tuning.}

\begin{figure}[h]
\centering{}
\includegraphics[width=5.2in]{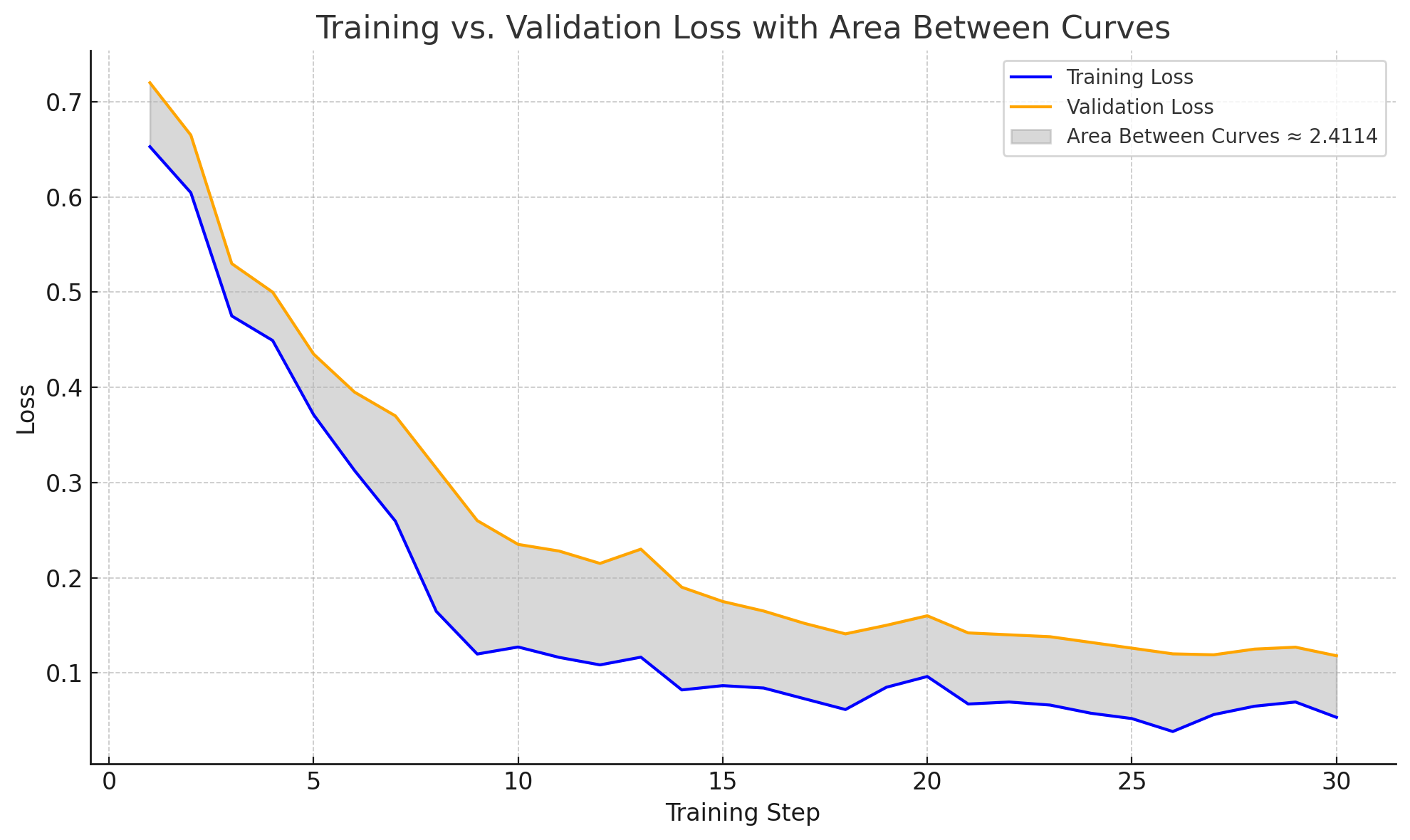}
\vspace{-0.1in}
\DeclareGraphicsExtensions.
\caption{Training loss and validation loss during fine-tuning of the Llama-3.2-11B LLM.}
\label{unsloth-tranning-validation-loss}
\end{figure}

\begin{figure}[h]
\centering{}
\includegraphics[width=5.2in]{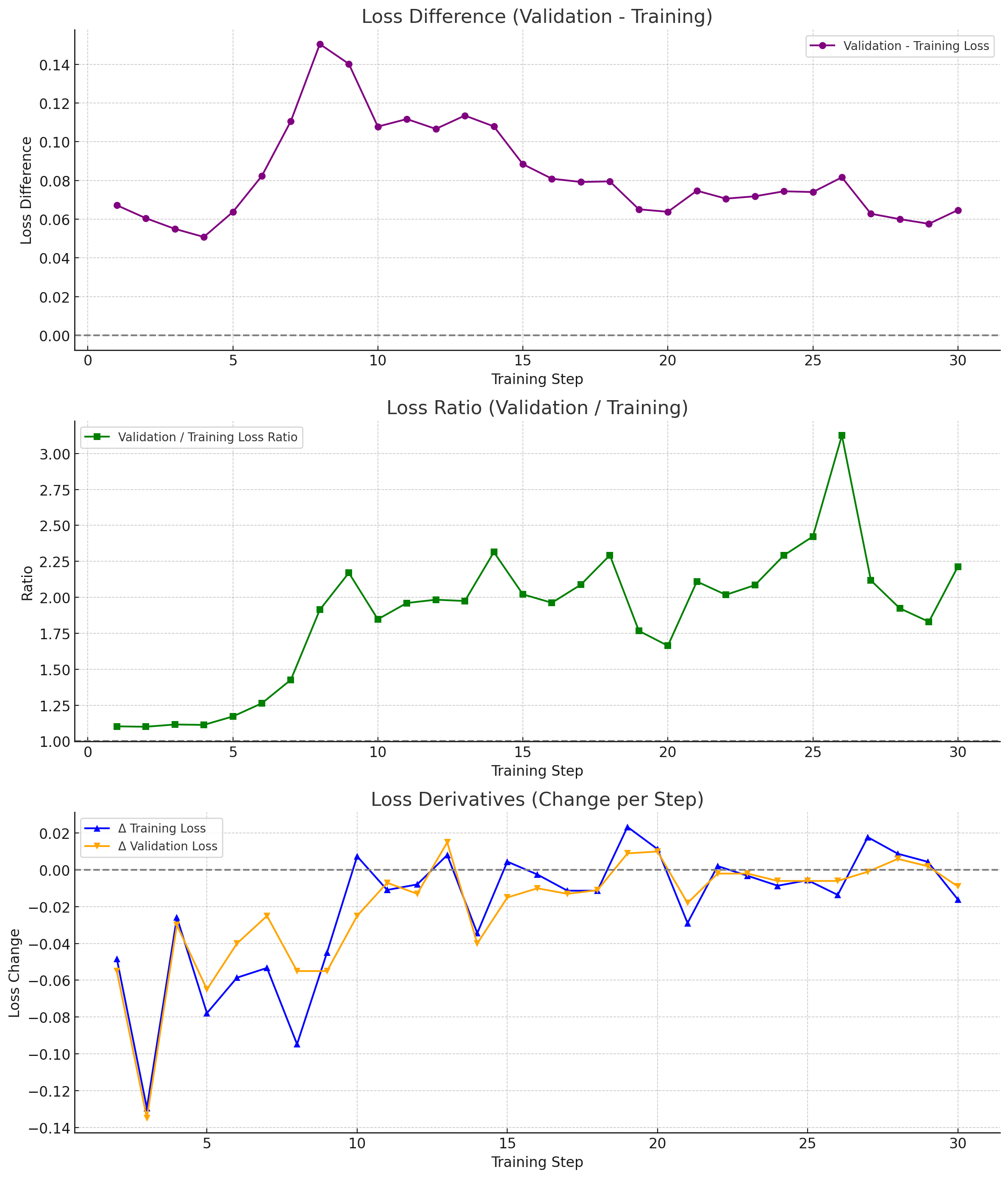}
\vspace{-0.1in}
\DeclareGraphicsExtensions.
\caption{Ratio of training to validation loss during the fine-tuning of the Llama-3.2-11B-Vision-Instruct VLM.}
\label{unsloth-loss-ratio}
\end{figure}

In this evaluation, we first assessed the training and validation loss during the fine-tuning process of the VLMs for diagram-based threat modeling. These metrics, visualized in Figure~\ref{unsloth-tranning-validation-loss}, demonstrate the models’ progressive learning across training steps. Furthermore, Figure~\ref{unsloth-loss-ratio} captures multiple key training dynamics, including the loss difference, loss ratio, and loss derivatives over training steps, offering valuable insights into the model’s convergence behavior and generalization performance. The consistently positive loss difference (validation loss exceeding training loss) suggests signs of overfitting, especially at steps with noticeable spikes. The loss ratio, ranging from 1.0 to 3.0, highlights varying degrees of generalization, where a lower ratio reflects better alignment between training and validation performance. Additionally, the loss derivatives reveal rapid initial improvements followed by smaller, oscillating changes, indicating stabilization or saturation in the learning process. 

\begin{table*}[!htb]\centering
\caption{Comparison of Software Development Methodologies}
\label{sdm-comparison}
\begin{adjustbox}{width=1\textwidth}
\begin{tabular}{lccccccc}
\toprule
\thead{Methodology} &
\thead{Team\\Collaboration} &
\thead{Iterative\\Approach} &
\thead{Project\\Planning} &
\thead{Adaptability\\to Change} &
\thead{Target\\for AI} &
\thead{Support for\\LLM Fine-Tuning} &
\thead{Applicability} \\

\midrule

Agentsway & High & Strongly Iterative & Adaptive and Lightweight & High & Yes & Yes & AI-agent–centric teams \\

Waterfall~\cite{waterfall} & Low & Sequential & Rigid, upfront & Low & No & No & Large, well-defined projects \\

Agile~\cite{agile-not-agile} & High & Iterative (Sprints) & Dynamic & High & Partial & No & Broad range of projects \\

Kanban~\cite{kanban} & High & Continuous flow & Visual and flexible & High & No & No & Streamlined operational tasks \\

Lean~\cite{lean} & High & Iterative & Minimal overhead & High & No & No & Process improvement initiatives \\

Spiral Model~\cite{spiral-model} & Moderate & Iterative cycles & Risk-driven & Moderate & No & No & High-risk, exploratory projects \\

XP~\cite{xp} & High & Iterative & Dynamic and adaptive & High & No & No & Small to medium-sized agile teams \\

Crystal~\cite{crystal} & High & Iterative & Lightweight & High & No & No & Human-centered agile projects \\

ShapeUp~\cite{shapeup} & High & Iterative cycles & Lightweight and scoped & High & No & No & Product-focused development teams \\

\bottomrule
\end{tabular}
\end{adjustbox}
\end{table*}

\section{Related Works}

Over the decades, software engineering has evolved through various development methodologies, each offering distinct approaches to managing complexity, uncertainty, and collaboration. Traditional models such as Waterfall~\cite{waterfall} and the Spiral Model~\cite{lean} emphasize structured progression, documentation, and risk management. Waterfall’s sequential design suits projects with stable requirements but limits adaptability, while the Spiral Model introduces iterative refinement and risk analysis to accommodate uncertainty. Lean Development~\cite{lean} extended this thinking by focusing on waste reduction, continuous improvement, and maximizing customer value through streamlined processes.

The rise of Agile methodologies~\cite{agile-not-agile, agile3-information-journal} marked a paradigm shift toward flexibility, collaboration, and iterative delivery. Frameworks such as Scrum, Kanban~\cite{kanban}, and Extreme Programming (XP)~\cite{xp} emphasize adaptive planning, customer involvement, and rapid feedback cycles. Agile methods encourage short, incremental iterations, making them highly responsive to change, while Kanban’s visualization techniques improve task flow and efficiency. XP contributes engineering rigor through practices like test-driven development and pair programming, ensuring both agility and code quality.

More recently, hybrid and human-centered methodologies such as Crystal Methods~\cite{crystal} and ShapeUp~\cite{shapeup} have emerged to balance autonomy with structured planning. Crystal adapts process intensity to project scale and criticality, emphasizing communication and team dynamics. ShapeUp, developed by Basecamp, introduces a “shaping” phase before implementation, promoting clarity of scope and outcome-driven work cycles. While these methodologies improved team productivity and adaptability, none explicitly address the growing role of autonomous AI agents in software creation—highlighting the need for new paradigms like Agentsway, designed to integrate planning, prompting, and learning within AI-agent–centric development environments.

\section{Conclusion and Future Works}
\label{sec:conclusion}

The rapid advancement of Agentic AI is redefining the foundations of software engineering. Traditional development methodologies, built around human-centric collaboration and manual iteration, are no longer sufficient to manage the autonomy, speed, and scale introduced by intelligent AI agents. This paper has presented ``Agentsway", a novel software development methodology that formalizes the interaction between humans and AI agents across the entire software lifecycle. Agentsway introduces a structured framework in which a human orchestrator collaborates with specialized agents—responsible for planning, prompting, coding, testing, and fine-tuning—to deliver software through a privacy-conscious and continuously learning process. The methodology’s design emphasizes explainability, responsibility, governance, and data privacy, ensuring that the integration of AI agents into development workflows remains both transparent and compliant with organizational and regulatory requirements. Through the legal case handling automation use case, we demonstrated the practical applicability of Agentsway methodology. The evaluation results confirmed that agentic orchestration improves planning coherence, prompt accuracy, and model reliability while maintaining human oversight. Fine-tuned LLMs and feedback-driven iteration further enhanced contextual reasoning and reduced error propagation across cycles. By uniting orchestration, autonomy, and privacy-by-design, Agentsway marks a significant step toward the next generation of AI-native software development methodologies. To the best of our knowledge, this is the first research initiative to introduce a dedicated software development methodology explicitly designed for AI agent–based teams. Ultimately, Agentsway lays the groundwork for sustainable, transparent, and ethically aligned software engineering in the era of intelligent, self-improving agents. In future work, we plan to extend the adoption of the Agentsway methodology across diverse domains—including non–software engineering fields such as legal, tourism, and cybersecurity to evaluate its scalability, interoperability, and domain adaptability.




\bibliographystyle{unsrt}
\bibliography{reference}

\end{document}